\documentclass[webpdf,modern,large]{oup-authoring-template}

\graphicspath{{figures/}}

\theoremstyle{thmstyleone}%
\theoremstyle{thmstyletwo}%
\theoremstyle{thmstylethree}%

\usepackage{csquotes}
\usepackage{tikz}
\usepackage{booktabs}
\usepackage{multicol}
\usepackage{multirow}
\usepackage{makecell}
\usepackage{colortbl}
\usepackage{threeparttable}
\usepackage{paralist}
\usepackage{xspace}
\usepackage{bookmark}
\usepackage{cleveref}
\usepackage{graphicx}
\usepackage[caption=false]{subfig}
\usepackage{hyperref}
\usepackage{xurl}

\begin{document}

\journaltitle{Journal of Cybersecurity}
\DOI{\href{https://doi.org/10.1093/cybsec/tyag026}{10.1093/cybsec/tyag026}}
\copyrightyear{2026}
\pubyear{2026}
\appnotes{Paper}

\firstpage{1}

\title[Experiencing Apple's Lockdown Mode]{Experiencing Apple's Lockdown Mode -- The Challenges of Providing Technology for At-Risk Users\thanks{This is a pre-copyedited, author-produced version
of an article accepted for publication in \emph{Journal of Cybersecurity}
following peer review. The version of record is available online at
\url{https://doi.org/10.1093/cybsec/tyag026}.}}
\author[1,$\ast$]{Benedikt Mader}
\author[1]{Christian Eichenmüller}
\author[1]{Gaston Pugliese}
\author[2]{Dennis Eckhardt}
\author[1]{Zinaida Benenson}

\authormark{Mader et al.}

\address[1]{\orgdiv{Department of Computer Science}, \orgname{Friedrich-Alexander-Universität Erlangen-Nürnberg}, \orgaddress{\street{Martenstr. 3}, \postcode{91058 Erlangen}, \country{Germany}}}
\address[2]{\orgdiv{Institute of Sociology}, \orgname{Friedrich-Alexander-Universität Erlangen-Nürnberg}, \orgaddress{\street{Fürther Str. 246c}, \postcode{90429 Nürnberg}, \country{Germany}}}

\corresp[$\ast$]{Corresponding author. \href{email:benedikt.mader@fau.de}{benedikt.mader@fau.de}}


\abstract{
Lockdown Mode, introduced in 2022 as an optional security hardening setting for Apple's operating systems, aims to protect users from ``some of the most sophisticated digital threats''. We present the first academic analysis of Lockdown Mode based on a three-month autoethnographic study of its everyday use. Our findings show that Lockdown Mode does not adhere to most principles proposed by Matthews et al. (2025) for technologies supporting prevention and monitoring of digital threats for at-risk users. Apple provides limited information about the underlying threat model and affected functionality, making it difficult for at-risk users to understand and evaluate the tool. Usability challenges further highlight the need for more granular controls, while the high volume of notifications offers little support for attack detection and instead contributes to user annoyance. Although we consider Lockdown Mode an important step toward improving security, we believe Apple should integrate principles for technology used by at-risk users more fully.
}



\maketitle

\section{Introduction}
\label{introduction}

Smartphones have become the most dominant kind of mobile devices~\citep{statista2023mobileDevicemarketShare}, driving over 51\% of global website traffic since 2019~\citep{statista2024mobileDeviceTraffic} and significantly influencing user behavior.
Precisely because of their general prevalence, however, smartphones have naturally become the target of cyber attacks. Due to the widespread use of smartphones among potential high-value targets in the corporate, governmental, and political sector, strong threat actors such as nation-state adversaries have recently come into focus within the mobile security landscape. Accordingly, sophisticated and expensive spyware such as Pegasus~\citep{triple2023marczak, pegasus2023dack} is being developed and sold to cover their needs. The murder of the journalist Jamal Khashoggi in 2018 highlights the serious consequences the infection of devices can have \citep{Kirchgaessner_2021}. Several members of his inner circle, including his fiancée, were targeted by Pegasus shortly before and after the assassination, which likely facilitated its successful execution.

At-risk users represent a diverse group of individuals who face an elevated threat of becoming victims of digital attacks, and who would experience disproportionate harm from such incidents~\citep{warford2022sok}. They include journalists~\citep{di2021we}, activists~\citep{albrecht2021collective}, academics~\citep{tanczer2020online}, refugees~\citep{simko2018computer}, quasi-public internet personas~\citep{samermit2023millions}, sex workers~\citep{mcdonald2021stressful}, and survivors of intimate partner violence~\citep{bellini2023digital}. To address these vulnerabilities, there exist both safety guides tailored to at-risk users~\citep{boyd2021understanding, rosenbloom_activists_2022} and technologies specifically designed for their protection~\citep{tadic_design_2023, mcgregor2015investigating}. However, the design of such technologies is complex, as the priorities of low-risk users, at-risk users, and developers often diverge~\citep{halpin_co-ordinating_2018}. A framework for identifying the requirements of technology designed for at-risk users was recently proposed by Matthews et al.~\citep{matthews_supporting_2025}.

\emph{Lockdown Mode}, released by Apple on September 12, 2022 for iOS~16, aims to be such a technology. 
Lockdown Mode is an optional hardening setting to safeguard at-risk users from very sophisticated attacks by limiting or disabling certain functionalities of the operating system and of system apps. 
Affected features include messaging, web browsing, FaceTime, and connecting to computers or wireless networks \citep{lockdown16ios,lockdown17ios,lockdown18ios,lockdown26ios}. Lockdown Mode can be enabled in the device settings of iOS (``Privacy \& Security''). However, Apple emphasizes that the activation of Lockdown Mode is not a common necessity, stating that ``most people are never targeted by attacks of this nature'' \citep{lockdown16ios}.
Yet, as we learned as part of an associated, yet unpublished research project with interviews of international democracy activists, Lockdown Mode might remain unknown among at-risk user populations. Since its release, modest publicity and discussions have largely been confined to specialist tech audiences. The limited coverage of Lockdown Mode includes a number of media articles~\citep{arstechnica2022Lockdown, techcrunch2022lockdown,wired:2024:lockdown,fake2023jamf,ljubuncic_lockdown_2023,eff_how_2024,larson_lockdown_2024,doffman_apple_2024,franceschi-bicchierai_apple_2026,cox_fbi_2026,schneier_iphone_2026,brodkin_fbi_2026}, posts~\citep{hackernews2022lockdown,reddit2022reviewOfLockdownMode,privacyGuides2023lockdown,privsec_consulting_apple_2024,reddit_is_2024,lucky-offer-1421_activated_2026}, and videos~\citep{petriIT2022, reneRitchie2022lockdown, techlore2023,billy_ellis_what_2026} in mostly tech-specific contexts. We are not aware of any published academic studies on Lockdown Mode.

We address this research gap by an exploratory study investigating 
the impact of Lockdown Mode on day-to-day life and its capabilities as a technology for at-risk users. 
As a result of an interdisciplinary collaboration 
between computer science and anthropology, 
we utilized autoethnography
to document the experiences 
of the first author while using an iPhone with Lockdown Mode 
over a period of three months.

A previous version of this paper was uploaded to arXiv \citep{mader_arxiv_2024}\footnote{\url{https://arxiv.org/abs/2411.13249v1}; the current version includes major revisions, e.g., integrating our findings into the Users States Framework and principles by Matthews et al. \citep{matthews_supporting_2025}.}.

\subsubsection*{Research Questions}

Our autoethnographic study of Lockdown Mode in iOS was guided by the following high-level research questions:

\begin{enumerate}
    \item \emph{What is the impact of Lockdown Mode on user experience in everyday use?}
    \item \emph{What does Lockdown Mode teach us about technologies for at-risk users?}
\end{enumerate}

\subsubsection*{Contributions}

The contributions of this paper are summarized as follows:

\begin{enumerate}
\item To the best of our knowledge, we conducted the first academic study on Apple's Lockdown Mode in iOS to investigate user experience during daily usage. 
Over a period of three months, the first author utilized autoethnography to document the hands-on experiences of using Lockdown Mode. Despite the subjective nature of autoethnography, and despite none of the authors being at-risk users at the time of writing, our observations can be extrapolated to achieve improvements for this user group.
\item We identified both conceptual and implementation issues in Lockdown Mode and analyzed it according to the User States Framework and design principles for at-risk user technology outlined by Matthews~et~al.~\citep{matthews_supporting_2025}:
\begin{enumerate}
\item Lockdown Mode does not incorporate most principles associated with the \emph{Prevention} phase. In particular, the lack of \emph{precise} information from Apple regarding the intended user group, the underlying threat model, and the features affected by Lockdown Mode make it difficult for users to properly assess and understand the tool, and to make an informed decision about whether to enable it. This approach to security, sometimes described as ``authoritarian''~\citep{adams1999users} or ``paternalistic''~\citep{dodier2017paternalistic}, has been shown to be harmful to both users and overall security.
\item Lockdown Mode also fails to adhere to the principles of the \emph{Monitoring} phase. It offers no protection against (accidental) data sharing and provides no warnings about actual threats. Instead, users are overwhelmed with numerous notifications that rarely correspond to meaningful security risks.
\end{enumerate}
\end{enumerate}

\subsubsection*{Outline} 

\Cref{fundamentals} provides an overview on at-risk users, technologies for at-risk users, and Lockdown Mode, including respective related work. 
The methodology of our autoethnographic study is presented in \Cref{methodology}. 
\Cref{results}, which is written from the perspective of the first author, 
covers their autoethnographic experiences using the Lockdown Mode. 
Our findings are discussed in \Cref{discussion}, before the paper is concluded in \Cref{conclusion}.

\section{Background and Related Work}
\label{fundamentals}

This section provides background information and relevant existing work to understand Lockdown Mode in the context of digital-safety research with at-risk users.

\subsection{Digital-Safety Research}
\label{subsec:digital-safety}

Digital-safety research describes a growing body of work addressing ``a person's or a group's state of security, privacy, safety, and autonomy, as it relates to their digital footprint'' \citep{bellini2024sok}. Special attention has been paid to the digital-safety needs of so-called ``at-risk users'', often also referred to as ``high-risk users''. These are users who face an elevated likelihood of a digital attack and/or would experience disproportionate harm from such an attack~\citep{warford2022sok}.

The realm of digital-safety research with at-risk users covers their digital security experiences~\citep{herbert2023digital, nagaraja2009snooping, scott2016security}, their collective information security in movement contexts~\citep{albrecht2021collective, boyd2021understanding, rosenbloom_activists_2022,tadic_design_2023, daffalla2021defensive, reisinger2023unified}, and their digital-safety needs as journalists~\citep{di2021we, mcgregor2015investigating, mcgregor2017weakest, tsui2021journalists}, academics~\citep{tanczer2020online}, refugees~\citep{simko2018computer}, quasi-public internet personas~\citep{samermit2023millions}, sex workers~\citep{mcdonald2021stressful}, or survivors of intimate partner violence~\citep{bellini2023digital, slupska2021threat, slupska2022aiding}. 
As this literature shows, it is important to identify differences in risk to avoid generalizing all at-risk users under a universal banner, as this could otherwise lead to faulty or insufficient harm mitigation~\citep{bellini2024sok}.

Recognizing the need for contextualization while upholding a common definition of at-risk users, Warford~et~al.~\citep{warford2022sok} undertook a comprehensive systematization study to establish a consolidated understanding of at-risk users. The study analyzed 95~papers across diverse populations, ultimately unveiling 10~overarching contextual risk factors that amplify digital-safety risks. These factors include societal elements such as political context and/or marginalization, relationship factors like dependence on a third party for digital support or having a relationship with an attacker, and personal circumstances such as prominence or access to sensitive resources.

\subsection{Technology for At-Risk Users}
\label{subsec:tech}

At-risk users, such as those participating in political protests, are commonly advised to disable data transmission, leave their phones at home, turn off biometric identification, use strong passwords, and rely on end-to-end encrypted messaging applications \citep{boyd2021understanding}. Nevertheless, many  users, particularly those newly exposed to digital threats from nation state actors, may lack the situational awareness or experience needed to understand these threats and implement appropriate protective measures, especially when their purpose is not clearly explained. For example, in a 2020 study, only 5 out of 50~Black Lives Matter activists expressed concerns about device vulnerabilities and hacking \citep{rosenbloom_activists_2022}, the very threats that Apple's Lockdown Mode seeks to mitigate.

Tadic et al. \citep{tadic_design_2023} conducted long-term research on activists in the Republika Srpska in Bosnia-Herzegovina. Their goal was to create a tool that would support the security and privacy of this group. They emphasized the importance of tailoring tools for at-risk users to their specific threat models, as, for example, determined through interviews. Since many at-risk users may prioritize more immediate concerns related to personal safety, activism, or daily survival, the tools must also focus on educating users and be as easy to use as possible, a point also highlighted by Halpin et al. \citep{halpin_co-ordinating_2018}. Furthermore, it is important for tools to be trustworthy, for example, by handling data locally and by publishing their source code. 

Slupska et al. \citep{slupska2022they} proposed ``participatory threat modeling'' as a way to incorporate the experiences of marginalized groups and to truly address their needs. As gender, race, or sexuality shape how cybersecurity is experienced, the inclusion of diverse sets of at-risk users within threat modeling practice is intended to counter oversimplified, top-down approaches to cybersecurity \citep{slupska2021participatory}.

Halpin et al. \citep{halpin_co-ordinating_2018} interviewed at-risk users, low-risk users, and developers about security properties of messaging protocols. A major challenge for technology providers lies in effectively incorporating contextual risk factors and delivering advanced protective measures that address the specific needs of at-risk users. 
More seasoned at-risk users typically possess a clear understanding of their threat model and often prioritize privacy-preserving features such as minimizing metadata collection to reduce surveillance risks and enable plausible deniability. However, according to Halpin et al. \citep{halpin_co-ordinating_2018}, achieving this goal is difficult because developers, at-risk users, and low-risk users often have markedly different needs and priorities.

One positive example of creating tools for at-risk users are the International Consortium of Investigative Journalists' (ICIJ) multiple collaboration platforms for analyzing the Panama Papers~\cite{mcgregor2017weakest}. By focusing on creating convenient and powerful tools, the ICIJ was able to keep the entire investigation, which involved nearly 400~journalists, secure and secret, even though many of the journalists had little experience with secure technologies.
ICIJ further facilitated this by creating strong security defaults, such as using PGP encryption for all communication, by providing good human support, and by fostering a shared sense of community to build trust in the system.

Matthews et al. \citep{matthews_supporting_2025} summarized their experience designing products used by at-risk users. They emphasized that being at-risk can be highly dynamic and diverse, and stress the importance of community building. They also proposed the User States Framework to better understand the needs and mindset of at-risk users. The framework categorizes the needs of at-risk users into four main stages: ``\emph{Prevention}, \emph{Monitoring}, \emph{Active Event} and \emph{Recovery}'' \citep{matthews_supporting_2025}. \emph{Prevention} consists of measures taken to minimize the threat of future events, such as activating Lockdown Mode in our case. \emph{Monitoring} means watching for signs that might indicate such an event. \emph{Active Event} refers to the high-stress situation during an actual attack, when the main goal is often to stop the attack as quickly as possible. This is typically followed by \emph{Recovery}, meaning the aftermath of an attack, such as evaluating how and why the attack occurred and mitigating the damage. These phases can be traversed in different orders, and different technologies are useful in different phases. Matthews~et~al.~\citep{matthews_supporting_2025} further expand on this by providing design principles for each stage that should be followed to create tools well-suited for at-risk users. These principles will be discussed in detail in \Cref{discussion}.

\subsection{Lockdown Mode}
\label{subsec:lockdown}

The Lockdown Mode introduced in iOS~16, which was released on September~12, 2022, is described by Apple as ``an optional, extreme protection that's designed for the very few individuals who, because of who they are or what they do, might be personally targeted by some of the most sophisticated digital threats'' \citep{lockdown16ios}. 
It is specifically tailored as a security feature for at-risk users facing heightened digital risks due to their identity or professional role. 
The primary goal of Lockdown Mode is to reduce the potential attack surface for highly targeted malware by restricting or disabling selected system and app functionalities.
Apple emphasizes that the activation of Lockdown Mode is not a common necessity, stating that ``most people are never targeted by attacks of this nature'' \citep{lockdown16ios}, 
and that functionality limitations may be experienced when using it. Lockdown Mode received an update with additional features in iOS~17 \citep{lockdown17ios} and remained unchanged in iOS~18 \citep{lockdown18ios}. The impact of later versions is discussed in \Cref{subsec:limitations}. \Cref{tab:lockdown_features} provides an overview of Lockdown Mode as advertised by Apple. During our study, we discovered additional restrictions that were not mentioned in Apple's information, such as restrictions on contact and destination sharing (see \Cref{subsec:usability}).

Lockdown Mode can be enabled in the device settings of iOS (``Privacy \& Security''). 
Users are provided with a brief explanation of Lockdown Mode and asked to confirm their decision, to ensure that the activation process is deliberate and informed.

\begin{table*}[tb]
\centering
\footnotesize
\caption{Lockdown Mode features in iOS 16, 17 and 18 according to Apple \citep{lockdown16ios,lockdown17ios,lockdown18ios}. Features added in iOS 17 are indicated in \emph{italics}. iOS 18 brought no new features for Lockdown Mode \citep{new2025apple}.}
\label{tab:lockdown_features}
\begin{tabular}{@{}p{3.2cm}@{\hspace{0.5\tabcolsep}}p{14.3cm}@{}}
\toprule
\textbf{Topics} & \textbf{Blocked/Unavailable Features (Literal Citations from \citep{lockdown16ios,lockdown17ios})} \\
\midrule
Messages & 
Most message attachment types are blocked, other than certain images, video, and audio. Some features, such as links and link previews, are unavailable. \\
\arrayrulecolor{black!20}\cmidrule{1-2}\arrayrulecolor{black}
Web Browsing & 
Certain complex web technologies are blocked, which might cause some websites to load more slowly or not operate correctly. In addition, web fonts might not be displayed, and images might be replaced with a missing image icon. \\
\arrayrulecolor{black!20}\cmidrule{1-2}\arrayrulecolor{black}
FaceTime & 
Incoming FaceTime calls are blocked unless you have previously called that person or contact \emph{Features such as SharePlay and Live Photos are unavailable}.  \\
\arrayrulecolor{black!20}\cmidrule{1-2}\arrayrulecolor{black}
Apple Services & 
Incoming invitations for Apple services, such as invitations to manage a home in the Home app, are blocked unless you have previously invited that person. \\
\arrayrulecolor{black!20}\cmidrule{1-2}\arrayrulecolor{black}
Shared Albums/\emph{Photos} & 
\emph{When photos are shared, location information is excluded.} Shared albums are removed from the Photos app, and new Shared Album invitations are blocked. You can still view these shared albums on other devices that do not have Lockdown Mode enabled. \\
\arrayrulecolor{black!20}\cmidrule{1-2}\arrayrulecolor{black}
Device Connections & 
To connect an iPhone to an accessory or another computer, the device needs to be unlocked. \\
\arrayrulecolor{black!20}\cmidrule{1-2}\arrayrulecolor{black}
\emph{Wireless Connectivity} & 
\emph{iPhone will not automatically join non-secure Wi-Fi networks and will disconnect from a non-secure Wi-Fi network when turning on Lockdown Mode. 2G cellular support is turned off.} \\
\arrayrulecolor{black!20}\cmidrule{1-2}\arrayrulecolor{black}
Configuration Profiles & 
Configuration profiles cannot be installed, and the iPhone cannot be enrolled in Mobile Device Management (MDM) or device supervision while in Lockdown Mode. \\
\bottomrule
\end{tabular}
\end{table*}

So far, research on Apple's Lockdown Mode is rather limited and we are not aware of any academic studies on either the usability and day-to-day use of Lockdown Mode or its technical aspects. 
Prior to Apple's implementation of Lockdown Mode in 2022, however, 
Gomez-Miralles and Arnedo-Moreno~\citep{lockup2015gomez} proposed a proof of concept for a security hardening of iOS in 2015, where they focused on ``[c]ontrolling the device's trust relationships'' \citep{lockup2015gomez}. 
Although most of the threats that Lockdown Mode attempts to address are outside the scope of their paper, the authors' solution to their particular problem (i.e., disabling certain features) appears to be similar to Apple's. 

There have been a limited number of test experiences evaluating Lockdown Mode. The most comprehensive assessment was conducted by the YouTube channel ``Techlore''~\citep{techlore2023}, which examined the impact of Lockdown Mode on daily use over a two-month period. Their findings revealed minimal friction during usage, and they consequently recommend Lockdown Mode for security- and privacy-conscious users.
A similar experience was shared by Newman~\citep{wired:2024:lockdown} who noted that iPhones in Lockdown Mode generally function as expected, although the restrictions can sometimes be quite severe. They note that it takes some getting used to the behavior of Lockdown Mode, as it is not intuitive from the start. They conclude that ``if you really need Lockdown Mode for your digital safety and personal protection, it's a workable alternative to throwing your phone in the ocean'' \citep{wired:2024:lockdown}. 

Regarding the technical capabilities of Lockdown Mode, Apple claims that as of March 2026 they ``are not aware of any successful mercenary spyware attacks against a Lockdown Mode-enabled Apple device'' \citep{franceschi-bicchierai_apple_2026}. In the case of US reporter Hannah Natanson, the FBI was unable to access her seized iPhone, on which Lockdown Mode was enabled \citep{cox_fbi_2026,schneier_iphone_2026,brodkin_fbi_2026}. While this cannot be conclusively attributed to Lockdown Mode, it may suggest anti-forensic capabilities.
Dack~\citep{pegasus2023dack} and Marczak~et~al.~\citep{triple2023marczak} indicated that Lockdown Mode can prevent and reveal the ``PWNYOURHOME'' attack, a new generation of Pegasus that exploits Apple's HomeKit. Marczak~\citep{marczak2023triangulation} has not yet reached a clear conclusion on whether Lockdown Mode can protect against ``TRIANGULATION'' vulnerabilities in iOS. Billy Ellis \citep{billy_ellis_what_2026} explained the technical fundamentals of Lockdown Mode and claimed that it removes the attack surface for most recent iOS browser exploits and blocks zero-click iMessage attacks from unknown contacts.
In contrast, Cryptee~\citep{ios2023cryptee} provided a proof of concept for detecting Lockdown Mode via browser fingerprinting, and Jamf Threat Labs~\citep{fake2023jamf} demonstrated the possibility of simulating active Lockdown Mode to users without providing any actual protection. 

Overall, there is still a lack of knowledge about Lockdown Mode, especially from an academic point of view and with a focus on everyday usability and user experience. Therefore, to the best of our knowledge, we have conducted the first academic study on this topic.

\section{Methodology}
\label{methodology}

\subsection{Autoethnography}
\label{methodology:rationale}

Autoethnography is a research method that involves documenting and analyzing personal experiences to enhance understanding of cultural phenomena \citep{ellis2004ethnographic, ellis2011autoethnography}. The term itself can be deconstructed into three components: ``auto'' signifies the utilization of personal narratives, ``ethno'' encompasses the exploration of cultural texts, experiences, beliefs, and practices, and ``graphy'' pertains to the descriptive and interpretive aspects of the method. It aims to handle the subjective beliefs of researchers transparently, without suggesting a false sense of objectivity. Autoethnography involves retrospectively and selectively narrating and critically analyzing significant events tied to one's cultural participation or identity.
When practicing autoethnography, researchers engage in prolonged fieldwork, actively participating and observing, and generating detailed \emph{field notes} that capture their encounters, assumptions, and potential biases \citep{adams2017autoethnograhy,eckhardt2023ethnografisches}. These notes serve as a raw and unfiltered record of the researcher's experiences. Following the fieldwork, the collected data is meticulously organized and structured into a coherent narrative suitable for outsiders.

In recent years, there has been a growing trend in the use of autoethnography within the HCI community \citep{o2014gaining,north2019imaginary,aagaard2023my,bala2023towards,lewis2023doodle,gaver2023living,vakeva_disorientation_24,lo_autoethographic_24,homewood2023self,wu_finding_24,chen_co_creating_24}, including security research \citep{turner2022hard,fassl2023can}.
Turner~et~al.~\citep{turner2022hard}, for instance, utilized the method to study problems associated with smart home security practices, while Fassl and Krombholz~\citep{fassl2023can} used autoethnography to understand the low adoption rate of authentication ceremonies. 

For our study of Lockdown Mode in iOS, we chose autoethnography as our primary research method, as it is particularly suitable when other methods would be unethical, dangerous, or potentially expose participants to risky investigation scenarios \citep{kaltenhauser_playing_24}. Since studies with at-risk users in the field are inherently risky to both study participants and researchers \citep{bellini2024sok}, an autoethnographic study of a hardening setting for Apple devices constitutes a useful, yet safe, approach. Given the exploratory nature of our research, our aim is to gain insights into Lockdown Mode in a safe environment. 

Studying the features of Lockdown Mode in a laboratory setting with at-risk users would not have adequately captured the spontaneous and highly personal ways in which smartphones are used, particularly in security-related contexts. Designing tasks that both feel realistic and comprehensively cover the changes introduced by Lockdown Mode would have been extremely challenging. Recruiting at-risk users for such a study would also have posed significant difficulties, as at-risk users often face more immediate concerns \citep{tadic_design_2023} than participating in such a time-intensive study, and participation might have exposed them to disproportionate risks.

Conducting an online survey with at-risk users about Lockdown Mode was similarly unsuitable. Based on our own observations and experiences from working with at-risk users, as well as the limited media coverage and online discussion surrounding Lockdown Mode at the time of investigation, we had strong reasons to suspect a low adoption rate. This would have made it difficult to investigate research questions beyond adoption itself.

Studying Lockdown Mode with non-at-risk users would have reduced the risks associated with working directly with at-risk populations. However, because non-at-risk users typically have different needs and priorities than at-risk users \citep{halpin_co-ordinating_2018}, findings from such a study would not have been directly transferable to the at-risk user context.

\subsection{Data Collection and Analysis}
\label{methodology:data}

\begin{figure*}[tb]
\includegraphics[width=\linewidth, alt = 
{Illustrated road map of our autoethnographic study visualised as Gantt chart covering the 7-month research period between June and December 2023, the version update of iOS 16 to iOS 17 in September, as well as the four consecutive study phases ``Journaling Practice'' (1 month), ``Refamiliarisation with iOS'' (1 month), ``Lockdown Mode'' (3 months), and ``Analysis'' (2 months). The first two phases helped the first author to getting used to journaling as a practice, before their experiences with Lockdown Mode were documented between August and October 2023. Afterwards in November 2023, audio recordings of relevant journal entries were created which helped to cluster the individual experiences and to identify overarching themes. Finally, the first author was interviewed by an co-author in December 2023 based on a first textual draft of their autoethnographic experiences with the aim to clarify unclear passages and for further reflection. Over almost the entire research period, the co-authors had weekly meetings between June and November 2023 in order to discuss observations and experiences of the first author with Lockdown Mode and to ensure the quality of the autoethnographic work.}
]{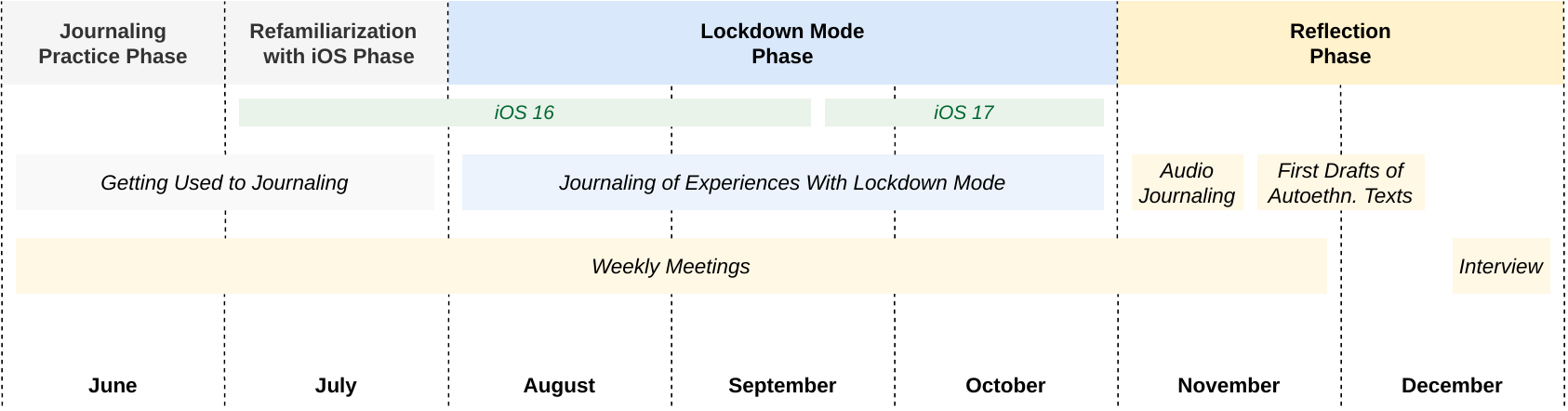}
\caption{Roadmap of the different study phases and the associated actions; months on the timeline correspond to the year 2023.}
\label{fig:phases}
\end{figure*}

The study was divided into four phases beginning in June 2023 (see \Cref{fig:phases}). During the first month, the first author used their current smartphone (OnePlus~6, Android/OxygenOS~11.1.2.2) to get used to the autoethnographic process. This was followed by another month of using the test device (iPhone~XR, iOS 16.5.1--17.0.3) without Lockdown Mode enabled to get a feel for ``normal'' iOS behavior. The iPhone was set up to mimic the first author's device, with a particular focus on using stock iOS apps and default settings. The main research phase lasted three months, during which the author used Lockdown Mode full time. Since iOS~17 was released in the middle of this period, including new Lockdown Mode features, both iOS~16 (starting with iOS~16.6) and iOS~17 (ending with iOS~17.0.3) could be investigated to a similar extent.

During these three practical phases, the first author collected their experiences on a daily basis. This was done through screenshots and evening journaling sessions. Journaling was conducted through a self-designed command-line script that prompted the user with predetermined questions and stored the answers in a digital repository, accessible to the co-authors. This process resulted in 127~screenshots and 415~journal entries. For a detailed overview of the journaling process, consult \Cref{app:journaling} in the appendix. During weekly team meetings, these first-order observations were discussed with the co-authors and supplemented by second-order observations. As the point of autoethnographic experience is to gain insights \emph{through a particular perspective}, these meetings allowed us to address the limits of the first author's perspective by reflecting on how they observe and what they might not yet have taken into account. Second-order observations and questions in team meetings have also assisted the first author in putting diffuse streams of experiences into words.

In the two-month analysis phase that followed, the first author reviewed the journal entries and screenshots. On this basis, they used audio recordings to reflect on relevant experiences and possible thematic clusters. The transcriptions of the recordings were summarized and clustered into overarching themes. As a final reflection loop, the first author was interviewed by a co-author in a one-hour session based on a first draft of \Cref{results} to clarify unclear passages and further reflect on the experiences. The interview was recorded, transcribed, and used to refine the textual elaboration.

\subsection{Persona}
\label{methodology:persona}

The first author is a security- and privacy-conscious computer scientist. The use of protective measures is not uncommon for them. Before embarking on the autoethnographic journey, they had no prior knowledge of Lockdown Mode and had not used iOS since 2018. Although they take their protection seriously, the first author lacks characteristics typically associated with at-risk users who might be advised to use Lockdown Mode (e.g., being a public figure or doing politically sensitive or controversial work).

However, the perspective of at-risk users was still incorporated through self-reflection and during team meetings, where we discussed how such users might react to specific situations that arose during the study. While the technical expertise of the first author may have led to experiences that differ from those of less technically skilled users, this expertise nonetheless strengthened the analysis by enabling a deeper understanding of the observed phenomena and supporting more nuanced and informed observations \citep{fassl2023can}.

\subsection{Ethical Considerations}
\label{subsec:ethical}

Ethical considerations were made before and during our study. As discussed in \Cref{methodology:rationale}, we actively decided against including real at-risk users to minimize risk as much at possible. Since Lockdown Mode is supposedly intended for users who face adversaries with powerful cyber capabilities, such as state agencies or actors with comparable resources, it is highly unlikely that the first author was actually at-risk while conducting the autoethnographic study.

Further, we needed to take precautions to protect the privacy of the first author and their social group, 
as smartphone usage touches on several deeply personal areas of life. To minimize invasiveness for the first author, we decided not to collect browser history. Also, the first author always had the option of omitting sensitive or embarrassing topics from their journal. However, this option was rarely used.

To protect their social group, the first author pseudonymized all mentions of other people who were not part of the study. The real identities of such individuals were not revealed to the co-authors. Screenshots containing real names or other identifiable information were censored, or not taken at all, if the content was deemed sensitive.

\section{Autoethnographic Experiences with Apple's Lockdown Mode}
\label{results}

This section is written from the point of view of the first author. Their perspective is presented in the first person, including excerpts from their journal and audio recordings. Spelling and punctuation errors in quoted journal entries were corrected to improve readability.

\subsection{``Trust Me, Bro'' — Navigating Apple's Information Void}
\label{subsec:communication}

As I prepared for the study and tried to enable Lockdown Mode in the device settings, I was immediately confronted with Apple's approach to communication. In the settings, I was presented with an overview of what features might be affected and the warning that ``apps, websites and features [will be] strictly limited for security'' and that ``some experiences [will be] completely unavailable''. This seemed to work well for a short introduction. However, \emph{``it doesn't really go into details what concretely works and doesn't work'' (Nov 09, Audio)}. Unfortunately, I could not find any additional documentation by Apple except for a slightly extended version of the claims during activation \citep{lockdown16ios,lockdown17ios} (see \Cref{tab:lockdown_features}). Furthermore, this list does not seem to be extensive. For example, the blocking of destination sharing in Maps \emph{``wasn't mentioned with a single point by Apple, as far as I have seen'' (Nov 14, Audio)}. 

Another matter that I found rather vague was the threat model. 
Apple states that ``Lockdown Mode helps [to] protect devices against 
extremely rare and highly sophisticated cyber attacks'' \citep{lockdown16ios}. 
However, there is no concrete overview of the possible attackers, their capabilities, the attacks, 
and how the features in Lockdown Mode help against them.

\begin{quotation}
\itshape 
I think Apple often has an attitude of `trust me bro, we are doing it the right way'. 
But you aren't able to evaluate it for yourself. (14 Nov, Audio)
\end{quotation}

If I were an at-risk user, I imagine I would find this very problematic because I would need to know exactly how I am protected by a feature like Lockdown Mode. Otherwise, I will not be able to assess where Lockdown Mode provides protection and where I may need to take further steps. So a lack of knowledge can lead to potential security risks. Also as a researcher,
it is difficult to evaluate the effectiveness of Lockdown Mode or to suggest improvements.

\begin{figure*}[htbp]
    \centering
    \begin{minipage}{0.33\linewidth}
        \centering
        \subfloat[Annotated missing website elements (Aug 20)]{\includegraphics[width=0.9\linewidth,
        alt = {Four screenshots illustrating the behavior of Lockdown Mode. The first screenshot shows a website for a hiking trail where the icons, images, and map are removed by Lockdown Mode.}
        ]{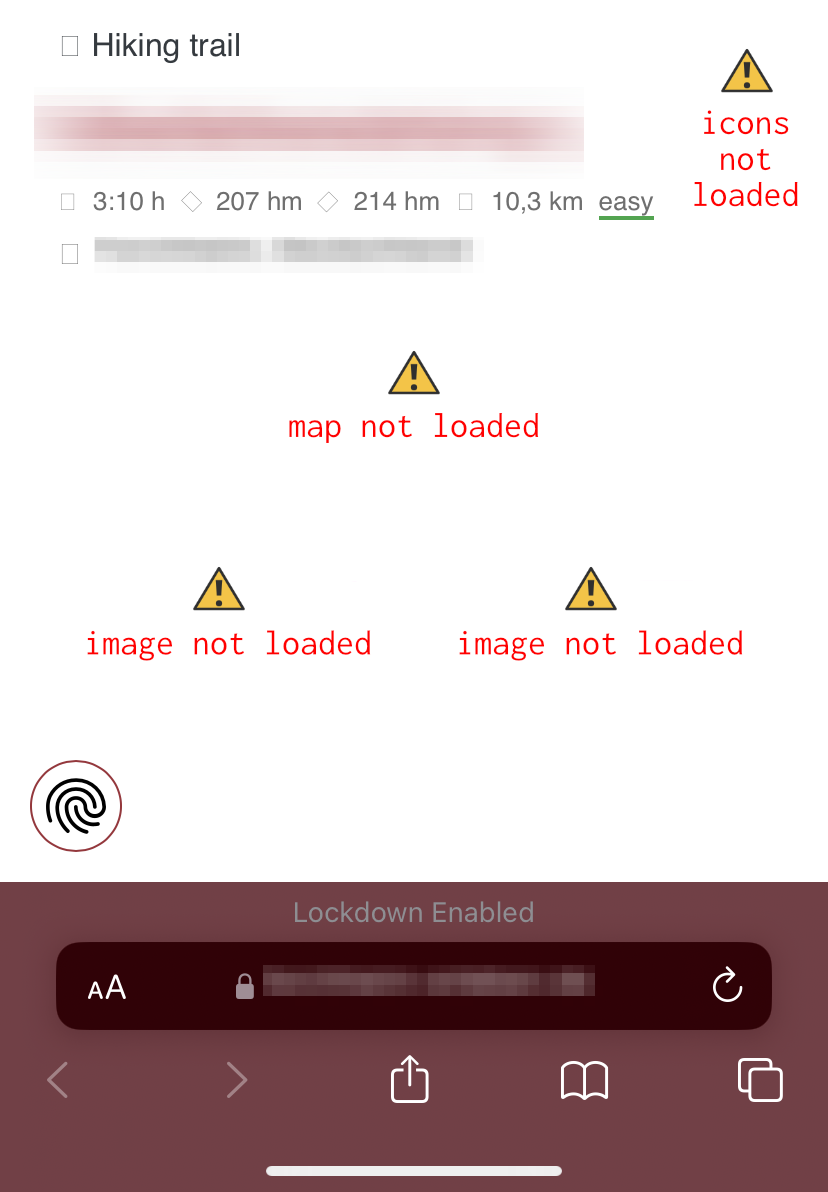}\label{fig:hiking}}\\
    \end{minipage}
    \begin{minipage}{0.33\linewidth}
        \centering
        \subfloat[Excluding website from Lockdown Mode (Aug~01)]{\includegraphics[width=0.7\linewidth,
        alt = {The second screenshot shows the notification for excluding websites from Lockdown Mode. Its text can be found in in \Cref{subsec:usability}.}]{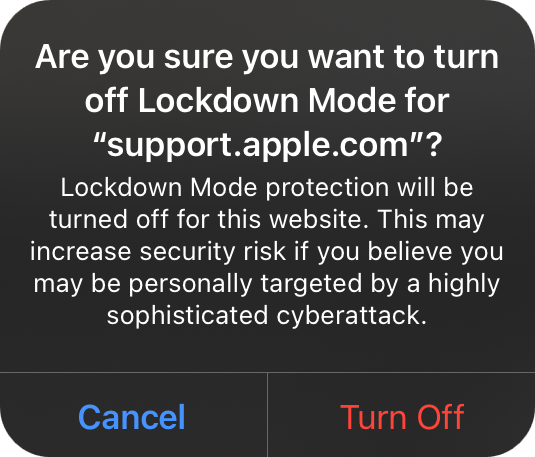}\label{fig:website}}\\
        \subfloat[Unable to open received files in iMessage (Oct 25)]{\includegraphics[width=0.7\linewidth,
        alt = {The third screenshot shows the notification when trying to open a locked iMessage file. It reads, "Cannot open message in Lockdown Mode --- This message cannot be opened in Lockdown Mode. You can turn off Lockdown Mode in Settings."}
        ]{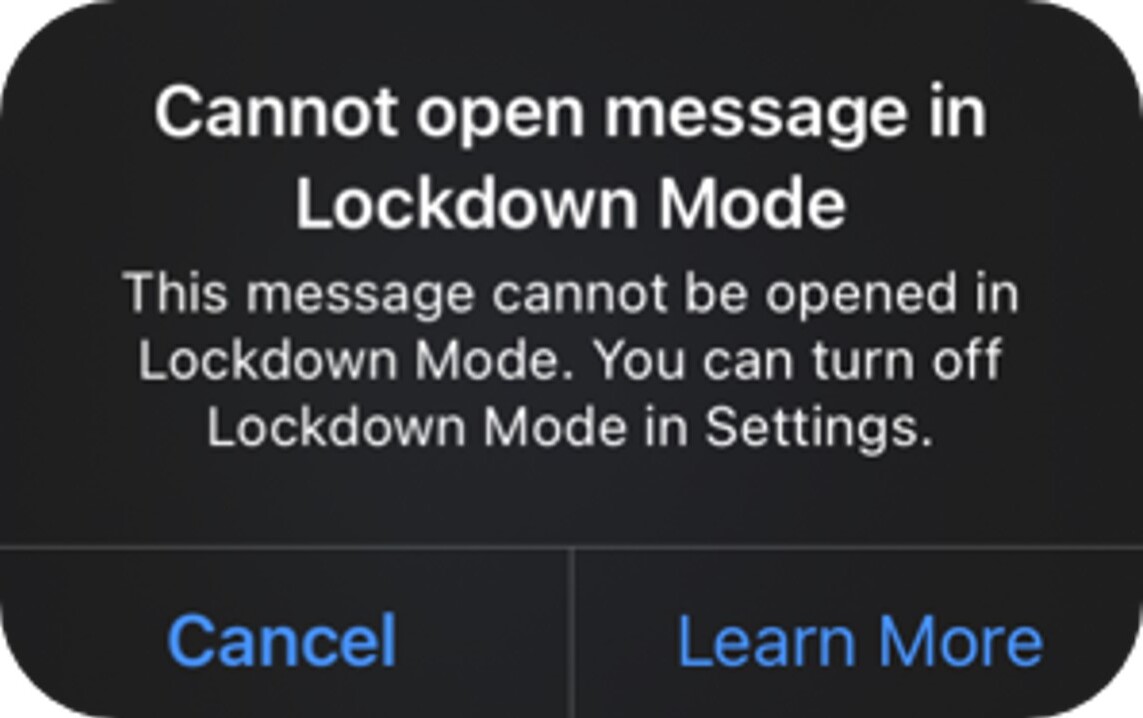}\label{fig:files}}\\
    \end{minipage} 
    \begin{minipage}{0.33\linewidth}
    \centering
        \subfloat[Lockdown Mode specific notifications (Oct 31)]{\includegraphics[width=0.9\linewidth,
        alt = {The fourth screenshot illustrates the overflow of notifications that can occur in the iOS ``Notification Center'' when notifications related to the Lockdown Mode accumulate over time. The notifications include blocking incoming FaceTime calls, blocking contact sharing, and multiple notifications indicating that Screen Time isn't being shared.}]{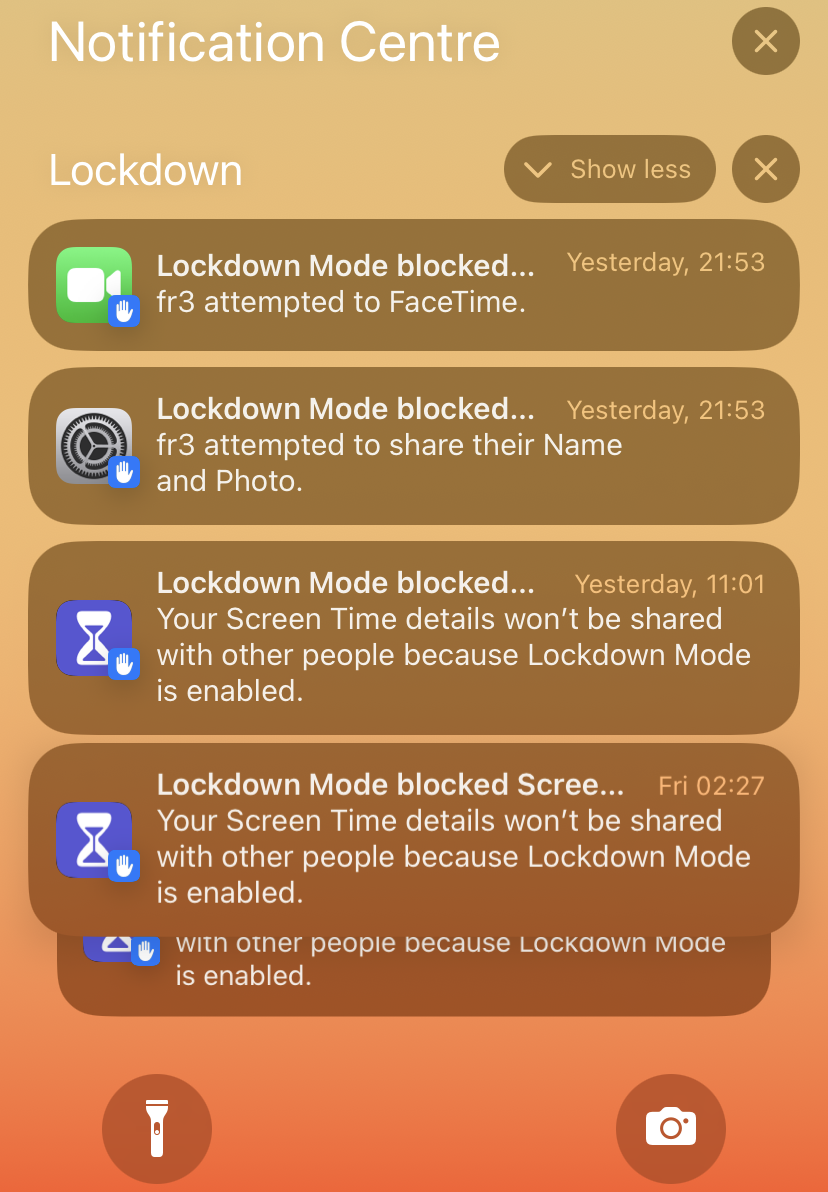}\label{fig:warning}}\\
    \end{minipage} 
    \caption{Screenshots of blocked functionality, warnings, and notifications that were were introduced by Lockdown Mode.}\label{fig:main}
\end{figure*}

\subsection{Encounters with Mysterious Design Choices}
\label{subsec:usability}

On several occasions, I was unable to use features of my device that I wanted to use. One memorable incident occurred during a hiking trip with a friend where the map showing the route was removed from the website (see \Cref{fig:hiking}). This meant that I had to delegate navigating to them, \emph{``but alone I would
have been helpless without deactivating Lockdown
Mode for the website.'' (Aug 20, Journal)}. However, Apple strongly discourages excluding websites from Lockdown Mode with a warning that reads (see \Cref{fig:website}): ``Are you sure you want to turn off Lockdown Mode for [website]? Lockdown Mode protection will be turned off for this website. This may increase security risk if you believe you may be personally targeted by a highly sophisticated cyberattack.'' So, as at-risk users might likely decide, I opted not to proceed with the exclusion.

Another annoyance was the usage of iMessage. While it was no big hindrance for me that link previews were not being shown, URLs are also not displayed as hyperlinks, but as regular text. This means that I had to copy and paste links into the browser manually if I wanted to access them. 
It is \emph{``quite annoying'' (Nov 13, Audio)} if the link is in the middle of a paragraph of text since it cannot be copied individually. Meaning, it \emph{``is just more work. But whether this is supposed to generate a big security advantage, I don't know'' (Nov 13, Audio)}. Furthermore, many common file types, including PDF and DOCX, are blocked from being opened (see \Cref{fig:files}). There is no possibility of an exception, just the option to deactivate Lockdown Mode system-wide.

\begin{quotation}
   \itshape
   I could imagine that for a high-risk user, this could be a problem. 
   If they get an important file [\dots] that's relevant for them, 
   they have to turn off Lockdown Mode completely. And it might be questionable if they will return [to Lockdown Mode, I] would say, 
   because if this is something they have to do regularly, this is quite restricting. (Nov 13, Audio)
\end{quotation}

While iMessage itself is quite protected, I also encountered a situation where a communication with a friend simply switched to SMS (which is handled in the same app). This made me feel insecure as it drops the E2E encryption that iMessage normally provides \citep{privacy2023apple}. \emph{``I [found] it quite bizarre, that this is not warned more
clearly about. [\dots] Lockdown should prevent such mistakes'' (Oct 01, Journal)}.

Another limitation is that incoming FaceTime calls from contacts who have not previously facetimed with you will be blocked (see \Cref{fig:warning}). I believe this could be particularly challenging for some professions that are typically associated with being at-risk, such as journalists, who need to communicate with a wide range of regularly changing people.
There is no way to exclude certain calls without first calling the other person. I also discovered an inconsistent design where calls from the same person would show up as blocked when attempted from their Mac, 
but not at all when attempted from an iPhone. This \emph{``seemed more like a bug than a feature to me'' (Aug 23, Journal)}. 

At the time, it felt quite odd to me that the problems with iMessage and FaceTime 
\emph{``could be easily avoided by using Signal or something like that'' (Aug 16, Journal)} as only Apple's apps and features are restricted. 
Another puzzling design decision was the limitation of the contact and destination sharing feature. It blocks incoming sharing attempts, but not outgoing ones.
This is particularly extreme with contact information, as it is even enabled to automatically be sent by default. I assume such information disclosure might be problematic for at-risk users: \emph{``This could be much more dangerous. Perhaps a high-risk user might have given an alias to some contacts. Now it is possible to accidentally disclose your real name and photo to all your contacts'' (Sep 29, Journal)}.
Only later did I understand the reasoning behind these two design choices (see \Cref{subsec:education}). Namely, that Apple's apps have higher permissions, and that
Apple seems to be primarily concerned with zero-click exploits, not social attacks. This struck me as very inconsistent at the time, and may not be understood by other users with similar or lower levels of technical understanding.

\subsection{Notification Overflow}
\label{subsec:attention}

Lockdown Mode introduces a host of new notifications to iOS. Their high volume considerably degraded my user experience. 
The most frustrating example arose after joining my iCloud family group. 
Frequent notifications informed me that my Screen Time data would not be shared with them (see \Cref{fig:warning}). To me, showing this notification so many times appears poorly planned, as there is no benefit in this information after the first instance.

\begin{quotation}
    \itshape
    It's really annoying since there does not exist a method to turn it off. 
    My notifications for Screen Time are already off\dots Really annoying if this persists. 
    Really bad implementation. (Sep 07, Journal)
\end{quotation}

Another example is the aforementioned contact sharing feature. Per default, contact sharing is automatically triggered when any iPhone user makes a call, uses iMessage, or FaceTime. As incoming sharing attempts are blocked by Lockdown Mode, this resulted in me getting an additional notification every time someone wanted to contact me (see \Cref{fig:warning}). As I do not like to be distracted by my phone, this was extremely irritating. Overall, I am not of the opinion that increasing the notification load is a good strategy for increasing perceived security or improving user experience.

\subsection{``Could As Well Not Be Active'' — Tension Between Expectation and Experience}
\label{subsec:experience}

While many of the problems described above stand out as particularly negative, I noticed few big changes in everyday usage during most of the investigation period. Especially at the beginning, Lockdown Mode appeared as it \emph{``could as well not be active'' (Aug 05, Journal)}. 
I think, my expectations before the study were not entirely accurate. Based on Apple's claims and the notices that most apps give you when you first launch them, indicating that functionality might be limited, I expected \emph{``major things that don't work [and] that it's kind of frustrating'' (Nov 09, Audio)}. However, I felt conflicted about whether the lack of friction in using Lockdown Mode would be beneficial to at-risk users, or whether it would make them feel vulnerable:

\begin{quotation}
    \itshape
    On the one hand, a smooth experience is beneficial for the usability of the device. But users might also be unsatisfied that not much really changes [\dots] and don't really feel more secure therefore. (Aug 16, Journal)
\end{quotation}

A side effect of my expectations was that I sometimes attributed problems, especially those related to web browsing, to Lockdown Mode when they had nothing to do with it. A notable example was trying to buy a transit ticket, which failed without an error message. \emph{``This could have been Lockdown Mode, I didn't quite know (Nov 15, Audio)''}. It was only after trying again and talking to people with similar problems that I could be sure it was a problem with the app and not Lockdown Mode. However, due to Apple's minimal communication strategy (see \Cref{subsec:communication}), it is not obvious whether problems are related to Lockdown Mode or not. In my opinion, this leads to a negative user experience.

\section{Discussion}
\label{discussion}

The User States Framework proposed by Matthews et al.~\citep{matthews_supporting_2025} in 2025 provides a structured approach to understanding user behavior and needs in response to digital threats. Within this framework, the primary objective of Lockdown Mode is \emph{Prevention} as it restricts certain functionalities to reduce exposure to potential digital threats. However, once activated, Lockdown Mode predominantly operates in the \emph{Monitoring} phase where users who are at risk must remain vigilant for any threats targeting them. Notably, the phases of \emph{Active Event} and \emph{Recovery} do not appear to be directly addressed by Lockdown Mode. The subsequent sections will examine how well the design principles proposed by Matthews et al.~\citep{matthews_supporting_2025} are adhered to within the context of Lockdown Mode.

\subsection{Lockdown Mode as a Tool for Prevention}

\subsubsection{Provide Digital-Safety Education}

\label{subsec:education}

A critical shortcoming of Lockdown Mode is the absence of explicit information about its intended target group and of a comprehensive threat model. 
Lockdown Mode is advertised to users as offering ``extreme protection'' against ``some of the most sophisticated digital threats'' and on the other hand discouraged to activate it as ``[m]ost people are never targeted by attacks of this nature'' \citep{lockdown16ios,lockdown17ios}.
This means that people are expected to autonomously decide if they are potentially targeted by these very powerful attacks, with very little information from Apple. However, at-risk users might often lack sufficient technological skills to asses necessary security protections or have different priorities \citep{tadic_design_2023}. Activists, for instance, may worry about a multitude of safety concerns including the misuse of personal data or surveillance concerns, with only a minority having concerns about hacking and device vulnerabilities \citep{rosenbloom_activists_2022}. Providing actual education here would presumably be very helpful, so at-risk users can assess if Lockdown Mode is necessary for them.

From the information provided by Apple, it is not directly apparent against what types of threats Lockdown Mode protects.
It should be noted that Lockdown Mode primarily protects against certain technical attacks, not social attacks. This means that Lockdown Mode only covers a small subsection of security advice given to at-risk users like activists \citep{boyd2021understanding}. Common recommendations such as disabling biometrics, using strong passwords and end-to-end encrypted messaging apps, or being cautious when sharing information or photos, are not addressed by Lockdown Mode.

One problem is that not every feature or design decision is obvious from the start.
For instance, we only later understood the lack of restrictions on third-party messengers (see \Cref{subsec:usability}). 
We first believed that the restrictions imposed by Lockdown Mode are likely intended for system apps because these apps have known vulnerabilities that attackers could exploit. For example, features such as iMessage's link preview can enable zero-click attacks with automated executions of malicious payloads.
However, we now believe the focus on these apps is a result of their higher level of permissions.
This means that third-party apps are less appealing targets since they have a smaller attack surface, because they lack the privileged ``entitlements'' (i.e., permissions of the type ``com.apple.private.*'') \citep{appleEntitlements} that system apps have. From this viewpoint, Apple's design choice seems logical, but it is difficult for users to understand the reasons behind it. 

This seems to us an instance of the information policy that is called ``need-to-know'' in the military context, which Adams and Sasse describe in their seminal work~\citep{adams1999users}: The users are provided with the absolute minimum of information about a security measure, because more information is deemed to facilitate attacks. They additionally call this approach ``authoritarian'' and show that badly informed users develop ``wildly inaccurate'' mental models of threats and security measures, which in turn undermines security. Later, Dodier-Lazaro et al. \citep{dodier2017paternalistic} call this approach ``paternalistic'': Security goals and measures are set by security experts without considering user values.

This is a conceptional weakness of Lockdown Mode, which is independent of individual features. We believe that a more transparent information policy on the part of Apple would serve to empower at-risk users, enabling them to make well-informed decisions with regard to their security, and thus enhance the level of  acceptance by people who really would profit from Lockdown Mode. Currently, Lockdown Mode does not follow the principle of providing digital-safety education.

\subsubsection{Make Digital-Safety Options Easy to Find}
Lockdown Mode is difficult to discover for users who do not know where to look or who are unaware of its existence. If users are familiar with Lockdown Mode, they can find instructions provided on Apple's support page \citep{lockdown16ios,lockdown17ios,lockdown18ios}. To activate Lockdown Mode, users must open the Settings app, navigate to the Privacy \& Security tab, scroll to the bottom of the menu, open the Lockdown Mode sub-menu, scroll through a lengthy explanatory text, select the activation option, reconfirm their choice, enter their passcode, and restart their iPhone. Combined with the limited information Apple provides both inside and outside its operating systems, we expect that it is unlikely for at-risk users to discover Lockdown Mode unless they are already well-informed or technologically savvy. Consequently, Lockdown Mode does not adhere to the principle that digital-safety features should be easy to find.

\subsubsection{Design for Strong Digital Safety by Default}
\label{subsec:default}

With respect to the principle of strong digital-safety defaults, Lockdown Mode presents a mixed picture. Although iOS includes a range of built-in security features, the additional protections offered by Lockdown Mode are not enabled by default and therefore do not fully align with this principle. However, once Lockdown Mode is activated, all of its protections are enabled automatically, making it relatively straightforward for users to elevate their security posture.

It is difficult to determine whether Apple should include all features of Lockdown Mode as defaults. On the one hand, doing so would increase security for all users, including at-risk users, without requiring any manual activation. On the other hand, this approach would also expose all users to the usability challenges and limited functionality inherent in Lockdown Mode. This could be at odds with Apple's typical focus on providing an enjoyable user experience, and may be seen as a competitive disadvantage. 
The fact that Apple chose to introduce Lockdown Mode as a separate feature suggests that achieving this balance is not straightforward. 

\subsubsection{Provide Multiple Digital-Safety Options, Especially Ways to Delete}
This principle is not addressed by Lockdown Mode. The feature is essentially a one-size-fits-all solution that does not distinguish between different user risk profiles. Additionally, Lockdown Mode does not provide dedicated options for quickly or easily deleting potentially incriminating information. However, this may not align with the types of attacks that Lockdown Mode is designed to defend against.

\subsubsection{Provide Private Spaces on Shared Devices}
Neither Lockdown Mode nor iOS provides private spaces on shared devices. However, Lockdown Mode does not appear to be designed to protect against threats posed by individuals who already share the smartphone with the user. Consequently, this principle does not seem to be particularly relevant in the context of Lockdown Mode.

\subsubsection{Reduce the Size of Security Updates}
Apple does not provide an official list detailing the size of its security updates \citep{apple_security_2025}. Nevertheless, multiple online discussions highlight the difficulties users face when installing updates, particularly major updates that may also introduce new Lockdown Mode features, as these often require several gigabytes of free storage on the device \citep{ok-wind-1675_its_2024,andenshap_how_2024,theslowknife_how_2024}. Moreover, iOS does not offer users the option to install only security-related updates or Lockdown Mode-specific features. Consequently, Lockdown Mode does not appear to adhere to the principle of minimizing the size of security updates.

\subsubsection{Consider Supporting Older Devices and Operating Systems}
With respect to supporting older devices and operating systems, Lockdown Mode adheres well to this principle. Lockdown Mode has been available since iOS~16, which is still supported by Apple today \citep{wikipedia_ios_2025}. iOS~16 is compatible with iPhones beginning with the iPhone~8 \citep{belinski_ios_2026}. Since the iPhone~8 was released in September 2017~\citep{wikipedia_iphone_2025}, Lockdown Mode was initially accessible to users with devices less than five years old. At present, iPhones up to nine years old can run a version of iOS that includes Lockdown Mode, albeit not necessarily all its features. With iOS~16 or later installed on 94.4\% of devices \citep{belinski_ios_2026}, Lockdown Mode is therefore available to the overwhelming majority of iOS users.

\subsubsection{Let Users Trade Functionality for Digital Safety}

Blocking certain features can lead to several usability issues, as discussed in \Cref{subsec:usability}. Consequently, the principle of allowing users to trade functionality for digital safety appears particularly valuable in this context. Although Lockdown Mode permits the exclusion of specific websites, neither iMessage nor FaceTime offers comparable flexibility. For example, the restriction that only known contacts may call can be circumvented by calling the unknown contact first, but a dedicated option to exclude specific contacts would provide a more convenient and secure solution. Another issue arises from the blocking of most file types in iMessage without any mechanism to exempt selected files or contacts. Users cannot determine which files they are restricted from opening because filenames are hidden. The only workaround suggested by iOS is to disable Lockdown Mode altogether, which undermines the security benefits it aims to provide.

Granting users more granular control and allowing specific exceptions within Lockdown Mode could substantially improve the user experience. Where secure alternatives to blocked features exist, such as compatible third-party applications, iOS could also highlight these options. This would help users avoid the binary choice between remaining in Lockdown Mode and disabling it entirely. One example where the trade-off is implemented effectively is the blocked 2G connectivity, which can be individually disabled in the settings. However, since most features lack comparable configurability, Lockdown Mode cannot be said to adhere to the principle of letting users trade functionality for digital safety.

\subsection{Lockdown Mode as a Tool for Monitoring}

\subsubsection{Show the Reach of Info the User Shares Online}

Lockdown Mode does not have any features to address this principle.

\subsubsection{Make Data Sharing Transparent}

Data sharing is not made transparent to the user in Lockdown Mode.
Instead, it, for instance, solely blocks incoming data-sharing attempts and notifies the user, as reported in \Cref{subsec:usability}. In contrast, users are neither prevented from nor informed about sharing sensitive information themselves, for example, their contact details, which continue to be shared automatically by default, or their destination. This limitation is not immediately apparent. Only through discussions in our weekly meetings did we arrive at the critical realization that Lockdown Mode primarily targets technical attacks from external sources, while offering limited protection against social attacks that rely on voluntary user disclosure. Because users are not made aware of these protection gaps, the principle of ensuring transparency in data sharing is not addressed at all by Lockdown Mode.

\subsubsection{Show Digital-Safety Controls in Relevant Places}
\label{subsubsection:tension}

There are no security-related controls shown within individual apps but only within the system settings. As a result, enabling Lockdown Mode, and modifying the limited set of configurable options, is unnecessarily difficult.

Furthermore, once Lockdown Mode is enabled, users may find it challenging to determine whether it is actually active. The first author reported feeling uncertain about the activation status and wished for clearer indications that Lockdown Mode was turned on.
This highlights a general tension in the salience of security features: highly visible mechanisms risk annoyance (\Cref{subsec:attention}), whereas invisible ones risk a feeling of insecurity (\Cref{subsec:experience}).
This annoyance-anxiety tension is also described by Fassl et al. \citep{fassl_comparing_2022}, who find that many users of anti-stalkerware apps feel reassured by visible indicators of security such as a ``sticky icon in the status bar and a permanent notification''. For others, however, such reminders are disruptive and annoying; instead, a lack of negative experiences is correlated with increased trust in the app. Dourish et al. \citep{dourish_security_2004} argue for making security measures visible. Building on this, Distler et al. \citep{distler2019security} observed that users felt reassured when encryption was explicitly referenced in the interface of a voting system, while feeling less secure when the process was ``too quick and easy''. Stransky et al. \citep{stransky_limited_2021} support these findings in the context of E2E-messaging and note that even simple text disclosures can increase the ``perception of a tool's security, trust, and privacy'' while more complex indicators such as icons or animations provide only very limited additional benefits.

We therefore believe that system-wide explicit and well-calibrated visibility of Lockdown Mode's status, such as the “Lockdown Enabled'' indicator displayed in Safari (see \Cref{fig:hiking}), could help reassure users. On the other hand, users may have a false sense of security, may prefer to hide the presence of Lockdown Mode to avoid drawing suspicion \citep{halpin_co-ordinating_2018}, or may be annoyed as discussed above. It may therefore be advisable to allow users to choose their preferred level of visibility.

\subsubsection{Warn about Potential Digital-Safety Concerns}

Lockdown Mode does introduce a range of warnings about potential security concerns. However, many of these warnings refer to events that are harmless in the overwhelming majority of cases, such as another person sharing their contact information (see \Cref{subsec:usability}). 
Due to this high false-positive rate, the warnings were mostly experienced as annoying by the first author and do not appear to be effective in alerting users about genuine attacks. Moreover, Lockdown Mode does not include any functionality capable of detecting actual attacks and warning the user accordingly.

An improvement to Lockdown Mode would be to redesign the notification system so that each type of message (e.g., informing users that a contact's photo and name are not being shared) is shown only once, with occasional reminders that the restriction remains in effect. In contrast, some notifications, such as warnings displayed when users attempt to whitelist websites, seem designed to scare or pressure users into compliance, as criticized by Sasse~\citep{warningsscaring}, rather than to enhance their understanding of potential threats. It should also be possible for users to disable such notifications, either through a dedicated section of the notification settings or within the Lockdown Mode settings.
Apple's current warning system therefore does not effectively address this principle.

\subsection{Lockdown Mode's Ability in Supporting At-Risk Users}

Overall, Lockdown Mode does not appear to adhere to the majority of principles for the \emph{Prevention} and \emph{Monitoring} phases of at-risk user technology outlined by Matthews et al. \citep{matthews_supporting_2025}. The only principle it addresses well is the provision of support, and thus protection, for comparatively old devices and operating systems. 

A major improvement would be to better educate users by providing clearer information about the threat model as well as the capabilities and limitations of Lockdown Mode, and by presenting this information more prominently on Apple's website and within the operating system itself. Additionally, there may be opportunities to extend the capabilities of Lockdown Mode to address a broader range of threats, such as privacy-related attacks that exploit the accidental disclosure of information.

Furthermore, Lockdown Mode should offer more granular controls in relevant areas and revise the notification it sends. Notifications should focus on actual threats and avoid overwhelming users with irrelevant information while balancing the tension between annoyance and anxiety (see \Cref{subsubsection:tension}).

\subsection{Limitations}
\label{subsec:limitations}

As the first author was new to the methodology of autoethnography, in hindsight we would adapt parts of the journaling process. For instance, the journaling focused mainly on the observations made throughout the day and the reactions to them. However, it was not originally designed to systematically track the evolving understanding and thinking of the first author over time. While we highlight some changes in understanding between the Lockdown Mode Phase and the Reflection Phase or later write-up, it is not possible to accurately reconstruct all early expectations and mental models, or their changes over time, without introducing recall bias. Furthermore, it is possible that rare frustrating experiences, such as those described in \Cref{subsec:usability,subsec:attention}, stand out and are therefore more strongly remembered than mundane day-to-day usage, introducing some negativity bias.

The first author's technical background and absence of actual risk exposure shaped the study in specific ways. It enabled a systematic evaluation of Lockdown Mode's functional and design properties under real everyday conditions, including inconsistencies that require technical literacy to identify. What it cannot capture is the affective dimension of using Lockdown Mode under genuine threat: how restrictions feel when one is actually targeted, and how trade-offs between security and functionality are made when the consequences are real. These questions require different methods and participants, and remain open for future work.

Due to the complexity of iOS and different usage patterns, it was not feasible to include all possible features, apps and websites. Moreover, due to the continuous evolution of iOS, this study provides only a temporal snapshot, as features may change with future updates. The Lockdown Mode examined herein corresponds to the feature sets of iOS versions 16, 17, and 18 \citep{lockdown16ios, lockdown17ios, lockdown18ios}. The specific features covered are presented in \Cref{app:features}.
Any new features introduced in iOS~26 \citep{lockdown26ios} on September 15, 2025, and beyond (versions 19 through 25 do not exist), as well as potential unannounced minor changes between iOS~17 and iOS~18, fall outside the scope of this investigation. However, as of March~2026, only 59.8\% of iOS users have adopted iOS~26 \citep{belinski_ios_2026}, making our findings still relevant, particularly because many of our observations concern fundamental design decisions rather than specific features.

\subsection{Future Work}

Future work might explicitly focus on technical evaluations of Lockdown Mode. Tracing the past and future evolution of Lockdown Mode as a protective iOS feature through longitudinal studies could bring into view user needs and developer responses. Whether Lockdown Mode actually reaches its intended user group, and effectively delivers the kind of protection it promises, still needs to be thoroughly assessed.
Future work might also conduct a comparative analysis of similar systems, i.e., Android's Advanced Protection Mode introduced in 2025 \citep{google_advanced_2026,lee_advanced_2025} or security and privacy focused custom ROMs such as GrapheneOS \citep{grahpheneos_grapheneos_2026,de_rentiis_investigators_2026}.

While direct involvement of at-risk users is desirable, all such efforts must always center around \emph{their} needs and safety concerns \citep{bellini2024sok}. As protagonists and beneficiaries of this line of research, at-risk users could share their reasoning behind (non-)adoption and their experiences of Lockdown Mode or other systems through diary studies or qualitative dialogue. Such dialogue could range from unstructured narrative biographical interviews and semi-structured interviews to structured policy Delphi panels involving at-risk users and other stakeholders such as developers.
Research of this kind might benefit from cooperation with frontline organizations, such as the Electronic Frontier Foundation (EFF) and Citizen Lab.

\section{Conclusion}
\label{conclusion}
We conducted an autoethnographic study of the everyday impact of Apple's Lockdown Mode in iOS~16/17 to assess its suitability for at-risk users. We find that Lockdown Mode fails to uphold many of the principles outlined by Matthews et al.~\citep{matthews_supporting_2025} for tools designed with at-risk users in mind. Apple provides little education about relevant threats and offers minimal information about Lockdown Mode itself, making it difficult for at-risk users to discover, understand, and evaluate the tool. The usability challenges we encountered highlight the need for more granular controls of features and visibility of Lockdown Mode. The overwhelming volume of notifications does little to support attack monitoring and instead often becomes a source of annoyance.

Although future versions, such as iOS~26, may introduce new features, the fundamental design decisions underlying Lockdown Mode appear to remain unchanged. Further research on Lockdown Mode is still necessary, particularly regarding its technical properties and its real-world impact on at-risk users. While we regard Lockdown Mode as an important step toward improving security, we argue that Apple should more fully incorporate principles for at-risk user technology.

\begin{appendices}

\section{Journaling}
\label{app:journaling}

This section explains the journaling approach used for this project.
As using a smartphone is an activity that takes place throughout the day, we decided that the first author would write a journal every evening summarizing their experiences.
Occasionally, this was done during the day or in the morning after. 
We decided on a structured approach where each activity or cluster of activities would get its own entry. 
Each entry answered the following questions:

\begin{compactenum}
    \item What did you do? Think about the activity and the corresponding smartphone/app usage.
    \item At what time did this take place?
    \item Where did it take place?
    \item What was the social situation? Who was involved?
    \item Did you have any conflicts or problems?
    \item Do you have any additional thoughts or explanations?
\end{compactenum}

For each entry, the date, and the number of the activity for the day were recorded. A set of abbreviations was created to speed up the answering of questions 2--4. Questions 4--6 could be left blank if not required. An additional question called `metajournal' was used to reflect on the method itself.

\begin{figure*}[t]
\centering
\includegraphics[width=0.7\linewidth, alt =
{Screenshot of a command-line interface showing the execution of our journaling script written in Python using the alias Bash command `journaling`, including the individual questions asked to the autoethnographer to survey their experiences in a structured and repeatable manner.}
]{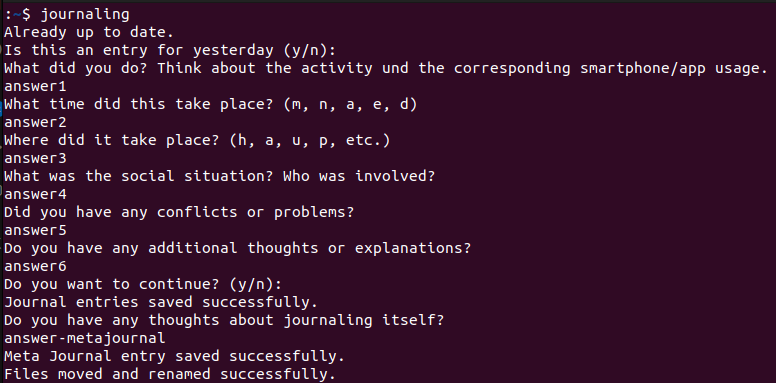}
\caption{Example of journaling script in action.}
\label{fig:script}
\end{figure*}

The questions were answered using a command-line script (see \Cref{fig:script}) that automatically saved the responses to a file and pushed them to GitLab, making the journaling process as low-barrier as possible while keeping the entries searchable and accessible to all authors.

In addition to written journal entries, the first author collected screenshots of notable situations.  An execution example of the full script is shown in \Cref{fig:script}.

Over the course of the study, the first author made 415~journal entries. 115 originated from the ``Journal Practice Phase'', 97 during the ``Refamiliarization with iOS Phase'', and 203 during the ``Lockdown Mode Phase''. This equates to approximately 2.7~entries per day over the whole 153-day study period and 2.2~entries a day during the 92~days of testing Lockdown Mode specifically. The journal entries were accompanied by 127~screenshots: 6 during the ``Journal Practice Phase'', 37 in the `Refamiliarization with iOS Phase', and 84 during the ``Lockdown Mode Phase'' (0.9 per day).

\section{Coverage of Lockdown Mode features}
\label{app:features}

\Cref{tab:testing} presents a list of features impacted by Lockdown Mode that were covered during the study, either because they were advertised by Apple \citep{lockdown16ios,lockdown17ios} or because they were encountered during the study and deemed relevant. 
The list was compiled in preparation for the autoethnographic phases and subsequently updated throughout the course of the study.

\begin{table}[!htb]
\centering
\begin{threeparttable}
\caption{List of specific iOS features covered during our autoethnographic study of using Lockdown Mode}\label{tab:testing}
\begin{tabular}{l c}
\hline
\multicolumn{1}{l}{\textbf{Features}} & \textbf{Coverage} \\ \hline
AirDrop                                 & covered           \\
AirPlay                                 & covered           \\
AirPods                                 & covered           \\
Attachments (iMessage)                  & covered           \\
Bluetooth                               & covered           \\
Configuration profiles                  & covered           \\
Contact sharing                         & covered           \\
Destination sharing                     & covered           \\
Device connections                      & covered           \\
FaceTime                                & covered           \\
Family sharing                          & covered           \\
FindMy                                  & covered           \\
iMessage                                & covered           \\
Insecure network                        & covered           \\
Links (iMessage)                        & covered           \\
Metadata (photos)                       & covered           \\
Photo classification (search function)  & covered           \\
Shared Albums                           & covered           \\
SMS                                     & covered           \\
Web browsing                            & covered           \\
Whitelist (web browsing)                & covered           \\
Wi-Fi hotspot                           & covered           \\ \hline
2G networks                             & not covered\tnote{a} \\
3rd party App Store                     & not covered\tnote{b} \\
Apple Cash                              & not covered\tnote{c} \\
HomeKit (Apple Services)                & not covered\tnote{d} \\ \hline
\end{tabular}
\begin{tablenotes}
\footnotesize
\item[a] Uncertain whether fallback to 2G ever occurred
\item[b] Not available during study period, only available since iOS 17.4 \citep{alternative2024apple}
\item[c] Only available in the US \citep{cash2024apple}
\item[d] Necessary smart home devices not available
\end{tablenotes}
\end{threeparttable}
\end{table}

\end{appendices}

\section*{Acknowledgments}
We would like to thank Julia Wunder and Felix Freiling for their valuable feedback, and Andreas Kurtz for his technical expertise. During the preparation of this work the first author used DeepL, ChatGPT, and Claude to improve language quality. After using these tools, the authors reviewed and edited all content as needed and take full responsibility for the content.

\section*{Competing interests}
No competing interest is declared.

\section*{Author contributions statement}
\noindent\textbf{Benedikt Mader:} Conceptualization (Equal), Data curation (Lead), Investigation (Lead), Methodology (Equal), Project administration (Equal), Visualization (Lead), Writing - original draft (Lead), Writing - review \& editing (Equal)

\noindent\textbf{Christian Eichenmüller:} Conceptualization (Equal), Investigation (Supporting), Methodology (Lead), Project administration (Equal), Supervision (Equal), Writing - original draft (Supporting), Writing - review \& editing (Equal)

\noindent\textbf{Gaston Pugliese:} Conceptualization (Equal), Investigation (Supporting), Methodology (Equal), Project administration (Equal), Supervision (Equal), Visualization (Equal), Writing - original draft (Supporting),  Writing - review \& editing (Equal)

\noindent\textbf{Dennis Eckhardt:} Methodology (Lead), Supervision (Supporting), Writing - review \& editing (Supporting)

\noindent\textbf{Zinaida Benenson:} Funding acquisition (Lead), Investigation (Supporting), Methodology (Supporting), Project administration (Equal), Supervision (Equal), Writing - original draft (Supporting), Writing - review \& editing (Equal)

\section*{Funding}
This work was supported by the Bavarian Ministry of Science and Arts as part of the project ``Security in Everyday Digitization'' (ForDaySec) and by Deutsche Forschungsgemeinschaft (DFG, German Research Foundation) as part of the Research Training Group 2475 ``Cybercrime and Forensic Computing'' (grant number 393541319/GRK2475/2-2024).

\bibliographystyle{unsrt}



\end{document}